# Integrating Color Histogram Analysis and Convolutional Neural Network for Skin Lesion Classification

M. A. Rasel[1], Sameem Abdul Kareem[1], Unaizah Obaidellah[1, *]

[1]Department of Artificial Intelligence, Faculty of Computer Science and Information Technology, Universiti Malaya, Kuala Lumpur, 50603, Malaysia
[*]Corresponding Author: Unaizah Obaidellah. Email: unaizah@um.edu.my

**Abstract**
The color of skin lesions is a crucial diagnostic feature for identifying malignant melanoma and other skin diseases. Typical colors associated with melanocytic lesions include tan, brown, black, red, white, and blue-gray. This study introduces a novel feature: the number of colors present in lesions, which can indicate the severity of skin diseases and help distinguish melanomas from benign lesions. We propose a color histogram analysis, a traditional image processing technique, to analyze the pixels of skin lesions from three publicly available datasets: PH2, ISIC2016, and Med-Node, which include dermoscopic and non-dermoscopic images. While the PH2 dataset contains ground truth about skin lesion colors, the ISIC2016 and Med-Node datasets lack such annotations; our algorithm establishes this ground truth using the color histogram analysis based on the PH2 dataset. We then design and train a 19-layer Convolutional Neural Network (CNN) with different skip connections of residual blocks to classify lesions into three categories based on the number of colors present. The DeepDream algorithm is utilized to visualize the learned features of different layers, and multiple configurations of the proposed CNN are tested, achieving the highest weighted F1-score of 75.00% on the test set. LIME is subsequently applied to identify the most important features influencing the model's decision-making. The findings demonstrate that the number of colors in lesions is a significant feature for describing skin conditions. The proposed CNN, particularly with three skip connections, shows strong potential for clinical application in diagnosing melanoma, supporting its use alongside traditional diagnostic methods

## 1. Introduction

Color variation represents a critical dermatological feature in diagnosing melanoma, a skin cancer originating from melanocytes, the cells responsible for skin pigmentation [1]. The number of colors present in a lesion is a key criterion for detecting melanoma, as specified by the ABCDE rules [2-5], the CASH algorithm [6], and the seven-point checklist [7-10]. These methods suggest that the presence of multiple colors within a skin lesion is often associated with malignant melanoma, with the total number of colors ranging from one to six. Melanomas exhibit a diverse color palette, including shades of brown, black, red, white, blue-gray, or combinations thereof [11-12]. Redness in a skin lesion typically indicates inflammation or increased blood flow to the area, which can be a sign of infection or malignancy [13]. Brown lesions may result from various causes, including moles (nevi), freckles, age spots (lentigines), or melanoma [14]. Light or dark brown areas may signify melanin production by melanocytes, which can be either normal or indicative of malignancy [15]. Black lesions often suggest melanoma, though some benign lesions, such as seborrheic keratoses, can also appear black [16][17]. White or pale lesions may result from conditions such as vitiligo, scarring, or areas of depigmentation following inflammation or injury [18], however, they can also indicate certain types of skin cancer or malignancy [18]. Blue-gray lesions might suggest a vascular component, such as a hemangioma or venous lake, or be indicative of melanoma [19]. Some melanomas can appear blue-gray due to the interaction of light with the pigmented cells [19]. Dermatologists often raise concern when a region within a lesion displays multiple colors or shades, presenting an irregular or patchy appearance indicative of potential malignancy [20]. Such lesions may manifest distinct colors in different areas, necessitating a thorough examination by trained dermatologists. **Fig. 1.** illustrates a skin lesion with various colors from the PH2 dataset [21].

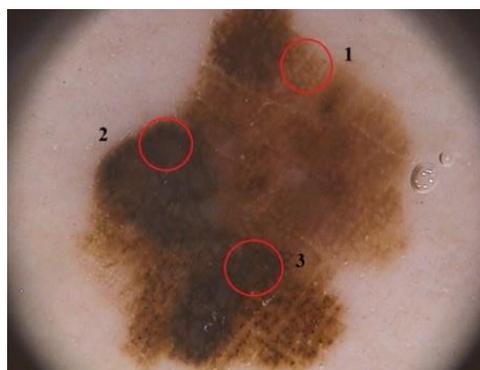

**Fig. 1.** Skin lesion image (IMD063.jpg) from the PH2 dataset [21] showing multiple colors in the region of interest (1 is light brown, 2 is dark brown, and 3 is black).

Furthermore, melanomas may exhibit colors that differ from the surrounding skin or change over time. It is important to emphasize that while color variation is a significant diagnostic factor, dermatologists consider it alongside other features when identifying melanoma [22]. Traditionally, recognizing diverse colors within a skin lesion relies on the discerning eyes of trained medical professionals, making the process prone to subjectivity [23]. The objective of this research is to develop an automated method for detecting color variations in dermoscopic images, leveraging both traditional image processing techniques and deep learning approaches. Specifically, this study aims to design and optimize a CNN architecture capable of accurately classifying skin lesions based on the number and types of colors present. By achieving this objective, the research seeks to enhance the efficiency and accuracy of melanoma diagnosis, ultimately improving patient outcomes and facilitating timely intervention. The methodological approach integrates traditional image processing techniques with a state-of-the-art deep learning model, specifically a 19-layer CNN designed to handle the complexity of color variation detection in skin lesion images. A distinguishing feature of the proposed CNN architecture is the use of residual blocks as skip connections in the hidden layers. These skip connections provide a way to directly pass the input of a layer to a deeper layer within the network, effectively skipping one or more intermediate layers, thereby allowing the network to capture intricate patterns and subtleties in the input data. In addition to the architectural design, extensive hyperparameter tuning is conducted to identify the most suitable configuration for color classification. Parameters such as learning rate, batch size, dropout rate, and optimizer are systematically adjusted and optimized to enhance the performance and robustness of the CNN model. By leveraging this comprehensive approach, the research aims to develop a highly accurate and reliable approach for automated color variation detection in skin lesion images, with a focus on achieving superior classification accuracy and ensuring that the model generalizes well across diverse datasets and clinical scenarios. **Fig. 2** shows the flowchart of the proposed approach.

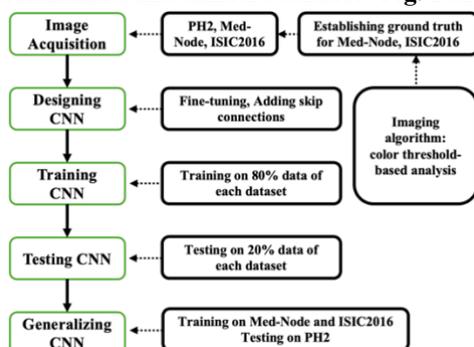

**Fig. 2.** The flowchart of the proposed approach to analyze and classify skin lesions based on the number of contained colors.

The major contributions of this research are as follows:
1. Developed an imaging algorithm to detect and differentiate colors within skin lesions effectively.
2. Established ground truth for non-annotated datasets, facilitating more accurate evaluation and validation of the model.
3. Designed a CNN model incorporating state-of-the-art techniques to classify skin lesions based on the number of colors present.
4. Validated the proposed model across diverse datasets to ensure its generalizability and effectiveness.

The rest of this paper is structured as follows: Section 2 reviews existing approaches for dermoscopic color analysis; Section 3 presents the skin lesion datasets, the proposed imaging algorithm for establishing ground truth in non-annotated datasets and describes the classifier architectures and implementation details; Section 4 shows the experimental results; Section 5 presents the discussions; and Section 6 provides concluding remarks and directions for future work.

## 2. Related Research Works

Dermatological color property has been the subject of investigation in various research works from the literature, holding significant importance for computer-aided diagnosis. However, limited research on lesion color detection can be attributed to several factors. Firstly, accurately capturing and quantifying skin lesion color consistently and objectively pose challenges due to variations in lighting conditions, imaging techniques, and subjective color perception. Moreover, the complexity of color analysis in the context of skin lesions, often featuring diverse colors and overlapping features, necessitates the integration of multiple clinical factors. Additionally, the lack of access to comprehensive and well-curated datasets of annotated dermoscopic images may impede the development and validation of color detection algorithms. These factors collectively contribute to the scarcity of research solely focused on lesion color detection. **Table 1** presents a systematic review of the dermatological color research. In the third column, "Analysis" means lesion analysis for dermoscopic features such as border, shape, and size; "Detection" means identifying different colors; and "Classification" means categorizing lesions based on the total number of colors.

**Table 1**
A review of skin lesion color analyzing approaches.

| Authors | Year | Aim: Color | Database | Techniques | Accuracy | Gap |
|---|---|---|---|---|---|---|
| Stanley et al. [24] | 2007 | Analysis, Detection | Atlas | Color Histogram, Machine Learning | 87.70% | Limited data, No classification, Semiautomatic |
| Nezhadian, & Rashidi [25] | 2017 | Analysis, Detection | ISIC | Active Counter Model, Machine Learning | 97.00% | Limited data, No classification, Semiautomatic |
| Almubarak et al. [26] | 2017 | Analysis, Detection | Atlas | Fuzzy Logic-based Color Histogram, Machine Learning | 92.60% | No classification, Semiautomatic |
| Rahman et al. [27] | 2017 | Analysis | Dermnet | RGB, HSV, YCbCr, KNN, SVM | N/A | No analysis of color's number |
| Mahmouei et al. [28] | 2019 | Analysis, Detection, Classification | Own | Color Enhancement, QuadTree Segmentation | 81.00% | Less number of colors, Semiautomatic |
| Moldovanu et al. [29] | 2021 | Analysis | PH2 Med-Node 7-point | kNN, RBFNN | 94.88% 94.71% 95.42% | No classification, Complex to implement, Focusing on disease recognition |
| Rasel et al. [30] | 2024 | Analysis, Detection, Classification | Augmented (PH2, ISIC), Derm7pt | Deep Convolutional Neural Network | 85.70% 95.05% 90.00% | Detected just only veils' color (blue-white) |

Stanley et al. [24] explored the use of relative color histogram analysis to discriminate skin lesions based on color features in dermoscopic images. This research examined different training set sizes and regions of the lesion for feature calculations. Experimental results demonstrated improved discrimination capability when focusing on the interior lesion region, achieving high recognition rates for malignant melanoma and dysplastic nevi. While the presented techniques, including relative color analysis and the determination of benign and malignant regions, hold potential applications for any set of benign and malignant lesion images, this research did not perform classification for the number of lesion colors. Nezhadian & Rashidi [25] introduced an algorithm for classifying dermoscopic images into malignant and benign lesions, with a specific focus on melanoma diagnosis. The algorithm utilized image segmentation, texture-based features, and color components for classification. Experimental results on the ISIC dataset demonstrated high accuracy using a support vector machine classifier. However, like the previous research, this study also did not consider the number of lesion colors, which is an important feature for diagnosing melanoma. Almubarak et al. [26] proposed a fuzzy logic-based color histogram analysis technique for distinguishing between benign and malignant lesions in dermoscopic images. The method utilized fuzzy sets and a fuzzy clustering ratio to quantify the skin lesion color feature. Experimental results on a dataset of 517 dermoscopic images demonstrated that the proposed approach achieved a recognition rate for melanomas, with a false-positive rate of approximately 13.5%. The results highlighted the significance of colors in the periphery of the lesion and the relevance of the fuzzy logic-based description for the clinical assessment of malignant melanoma. However, like the previous studies, color classification of the dermoscopic lesions was not conducted. Rahman et al. [27] proposed another skin lesion classification method to enhance existing diagnostic systems by employing advanced feature extraction and classification techniques on standard digital images. Initial preprocessing involves resizing, hair removal, noise filtering, and contrast enhancement. Segmentation in the Lab color plane using color thresholding enhances contrast, validated with an interactive segmentation tool. Higher-order statistics, histogram-based analyses, and image quality analysis are performed for feature extraction. Features are then used with machine learning algorithms to classify lesions into four classes (Melanoma, Basal Cell Carcinoma, Keratoacanthoma, and Squamous Cell Carcinoma). It's worth noting that, like other approaches, they did not conduct a separate lesion color analysis. Mahmouei et al. [28] introduced a QuadTree-based melanoma detection system, inspired by dermatologists' color perception. The system addressed challenges in clinical color assessment in dermoscopic images by employing color enhancement and automatic color identification techniques based on QuadTree segmentation. This approach accurately modeled expert color assessment, identifying more colors in melanoma compared to benign skin lesions and delineating the locations of melanoma colors. Despite these advancements, this research also did not classify or identify the number of colors in the lesions. Moldovanu et al. [29] proposed a novel approach for skin cancer recognition and classification by combining surface fractal dimension and color area features, which are classified using both a k-nearest neighbor algorithm with 5-fold cross-validation and a Radial Basis Function Neural Network (RBFNN). The surface fractal dimension is calculated via a 2D generalization of Higuchi's method, while a clustering method selects relevant color distributions in skin lesion images. This method, aimed at assisting dermatologists, was validated using images from various datasets. However, the study did not focus on individual dermoscopic color features based on the number of colors present; instead, it primarily focused on diagnosing melanoma. Rasel et al. [30] introduced a Deep CNN with 31 layers designed to classify skin lesions into two categories: blue-white veil and non-blue-white veil lesions, using three different datasets. In their approach, they carefully analyzed the color of lesions on a patch-by-patch basis to establish the ground truth regarding the presence of the veil. After training, the LIME algorithm was employed to visualize the most important features contributing to the model's decisions. Although the model achieved high accuracy, the research primarily focused on the blue-white veil color, leaving the analysis of other lesion colors as a direction for future work.

Although these approaches detected the RGB color of lesions and analyzed them to understand skin condition, most of the time they were individually applied to a single dataset, and the process of color classification (such as light brown, dark brown, black, red, white, and blue-gray) was absent to fit into ABCDE and CASH methods. Dermoscopic color classification in different colors is challenging due to several reasons: subjectivity and variability, complex color patterns, overlapping color features, atypical colors, and technical limitations. Besides, most of the detection methods are supervised learning techniques (without deep learning). Utilizing deep learning-based models can potentially provide a solution for this non-automatic process, enabling automatic color detection and classification. Additionally, some other research works guided to organize this research, such as melanoma identification using geometric features of super-pixels [31], lesion segmentation to validate melanocytic lesions using a color scoring method [32], and a hybrid CNN for region of interest segmentation [33].

## 3. Methods and Materials
### 3.1 Skin Lesion Image Acquisition

The skin lesion images used in this research were sourced from three distinct datasets. The first dataset, PH2 [21], was curated by the Dermatology Service of Hospital Pedro Hispano in Matosinhos, Portugal. It comprises 200 high-resolution skin lesion images (8-bit RGB color images with a resolution of 768x560 pixels), each annotated with detailed information on six specific lesion colors: white (19 occurrences), red (10 occurrences), light brown (139 occurrences), dark brown (156 occurrences), blue-gray (38 occurrences), and black (42 occurrences). These images were categorized into six classes based on the number of colors present. The second dataset, Med-Node [34], consists of 170 macroscopic multi-resolution images (with varying height-to-width ratios) sourced from the digital image archive of the Department of Dermatology at the University Medical Center Groningen. Unlike PH2, the Med-Node images lack annotations regarding skin lesion color. To establish the ground truth for this dataset, an imaging algorithm was developed to analyze pixel values in the lesion images individually (as detailed in the subsequent section). The third dataset, ISIC2016 [35], includes 1,279 dermoscopic images with resolutions ranging from 576x768 pixels to 2848x4288 pixels, of which 379 were selected for this study. This dataset, derived from the ISIC Archive, contains a representative mix of both malignant and benign skin lesions. Like Med-Node, the ISIC2016 images lack specific color annotations. Therefore, the same imaging algorithm used for Med-Node was applied to ISIC2016 to determine the presence and distribution of colors in the lesions.

### 3.2 Lesion Color Analysis

In this study, the PH2 skin lesion dataset provided annotations for six distinct lesion colors: red, light brown, dark brown, blue-gray, white, and black. To ensure consistency across the datasets used, it was necessary to establish the ground truth for the Med-Node and ISIC2016 datasets based on these same six lesion colors. However, the PH2 dataset did not include specific color space values for these colors, which presented a challenge in standardizing the classification process across multiple datasets. To address this, a method was employed based on visual assumption to define the RGB color ranges corresponding to the six lesion colors identified in the PH2 dataset. This approach aligns with the clinical practice where dermatologists and clinicians often make diagnostic decisions based on visual assessment of color scales [12]. By observing the lesion colors within the PH2 dataset, we established RGB ranges that closely matched the visual characteristics of each of the six colors, relying on standard RGB color values and visual interpretation to guide this process. **Table 2** shows the minimum and maximum RGB space values of those six lesion colors based on visual assumptions.

**Table 2**
The ranges of RGB space values of different skin lesion colors.

| Color's Name | Minimum | | | Maximum | | | Sample Color Patches |
|---|---|---|---|---|---|---|---|
| | Red | Green | Blue | Red | Green | Blue | |
| Light Brown | 170 | 95 | 27 | 190 | 105 | 30 | 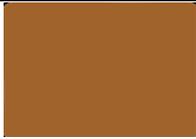 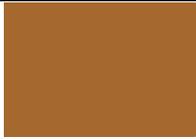 |
| Dark Brown | 89 | 59 | 29 | 110 | 73 | 36 | 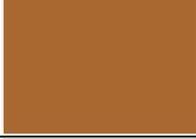 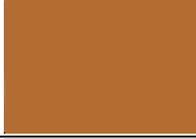 |
| Red | 155 | 25 | 33 | 255 | 80 | 60 | 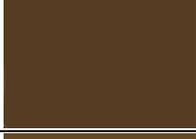 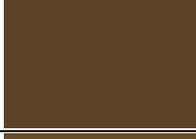 |
| Blue-Gray | 101 | 151 | 184 | 121 | 180 | 219 | 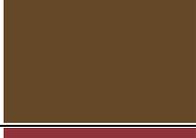 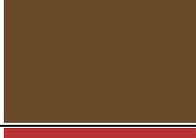 |
| White | 230 | 230 | 230 | 255 | 255 | 255 | 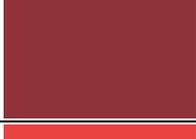 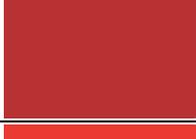 |

| | | | | | | | | |
|---|---|---|---|---|---|---|---|---|
| Black | 0 | 0 | 0 | 80 | 80 | 80 | 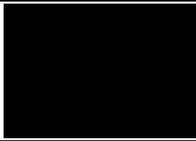 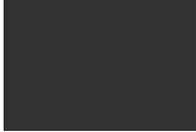 | 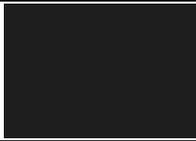 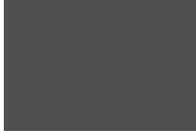 |

After defining the RGB ranges, an analysis of the PH2 dataset was conducted to validate our assumptions. We compared the RGB values of the lesions within PH2 against our predefined color scale and found that these values consistently matched the annotated ground truth provided by PH2. This confirmation indicated that our visual assumptions were accurate, and the RGB ranges we established were reliable representations of the six lesion colors. Having validated our color scale with the PH2 dataset, we applied the same RGB ranges to the Med-Node and ISIC2016 datasets to determine the presence of the six lesion colors in these datasets. **Table 3** shows the lesion color detection process in lesion images for PH2, Med-Node, and ISIC2016 datasets. This approach allowed us to maintain consistency in color-based classification across different datasets, ensuring that the ground truth for Med-Node and ISIC2016 was determined in a manner that is both visually and clinically relevant. By aligning this methodology with the visual assessment practices commonly used by dermatologists, we enhanced the robustness and applicability of this algorithm in real-world clinical scenarios.

**Table 3**
Detecting different colors on skin lesions of the PH2, Med-Node, and ISIC2016 datasets.

| Image | Light Brown | Dark Brown | Red | Blue-Gray | White | Black |
|---|---|---|---|---|---|---|
| 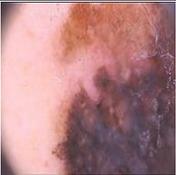 PH2 | 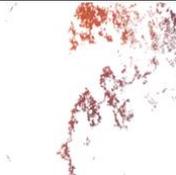 | 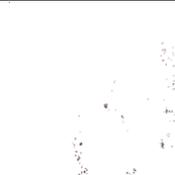 | 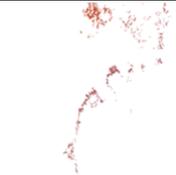 | 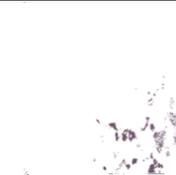 | 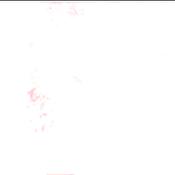 | Absent |
| 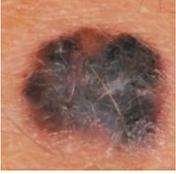 Med-node | 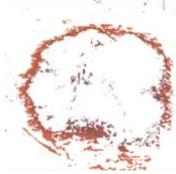 | 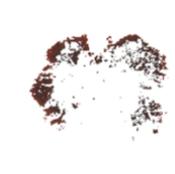 | 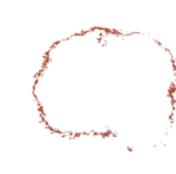 | 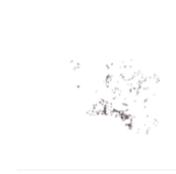 | Absent | 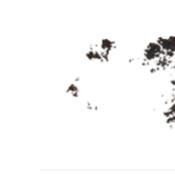 |
| 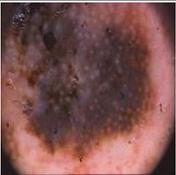 ISIC | 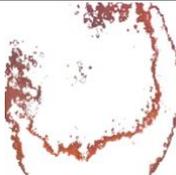 | 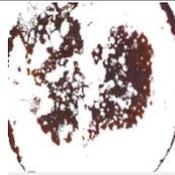 | 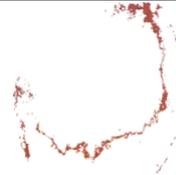 | 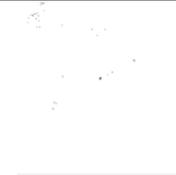 | Absent | 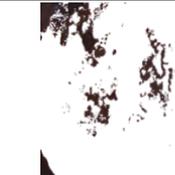 |
| 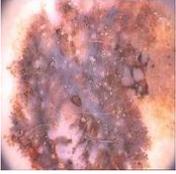 PH2 | 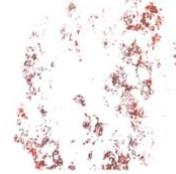 | Absent | 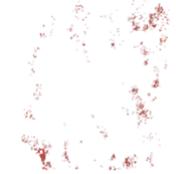 | 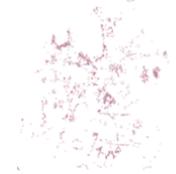 | 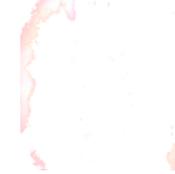 | Absent |
| 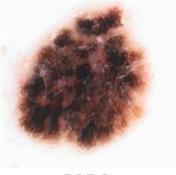 ISIC | 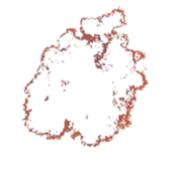 | 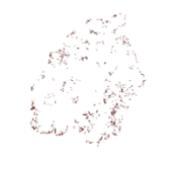 | 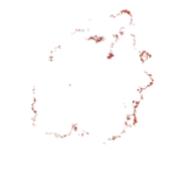 | Absent | 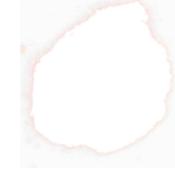 | 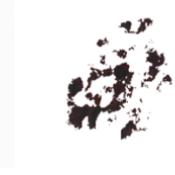 |

After identifying the presence of the six defined lesion colors (red, light brown, dark brown, blue-gray, white, and black) within the skin lesion images, a detailed analysis was conducted to minimize noise and ensure accurate color detection. Often, the detection of a specific color within a lesion image can be influenced by artifacts that do not represent the true color of the lesion. To address this, we implemented a process to refine the color detection by analyzing each color individually and verifying its relevance to the lesion.

Once a specific color was detected in an image, the image was divided into multiple patches, each with a patch size of 16x16 pixels. The rationale behind this approach is to assess the distribution and significance of the detected color across the lesion. If the number of patches containing the color exceeded five, we considered that the lesion genuinely contained that specific color. Conversely, if the color was present in five or fewer patches, we did not consider it significant and excluded it from further analysis. This threshold was established to differentiate between true lesion colors and incidental or artifact-induced color detections, thereby enhancing the reliability of our color classification.

For example, in the case of the skin lesion image ISIC_0000136.jpg from the ISIC2016 dataset, which was initially identified as containing five different colors, we analyzed the image separately for each color. During each analysis, the number of patches corresponding to the color in question was counted. As illustrated in **Table 4**, the image was analyzed twice for demonstration purposes, specifically for the light brown and black colors. Despite the image having a total of five colors as indicated in **Table 3**, the method demonstrated in **Table 4** was applied to the remaining three colors as well.

**Table 4**
Patch-by-patch analysis of detected colors on the skin lesion.

| Lesion Color | Lesion Image | Detected Color | Divided into Patches (16x16) | After Removing Background |
|---|---|---|---|---|
| **Light-Brown** | 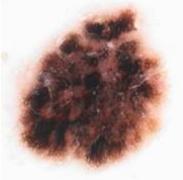 | 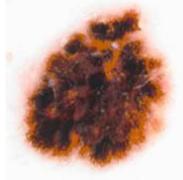 | 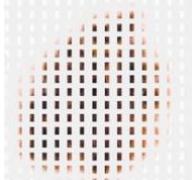 | 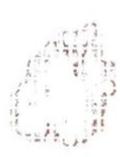 |
| **Black** | 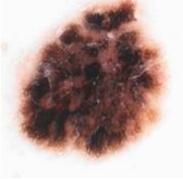 | 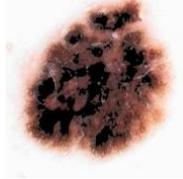 | 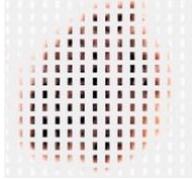 | 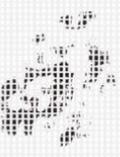 |

This patch-based analysis allowed us to systematically reduce noise and ensure that only significant color detections, likely to be clinically relevant, were included in the final classification. By refining the detection process in this manner, we increased the accuracy and robustness of the proposed imaging algorithm in distinguishing true lesion colors from irrelevant artifacts, thereby improving the reliability of the color-based classification of skin lesions across different datasets.

### 3.3 Evaluation Metrics

Evaluation metrics are necessary to provide quantitative measures that guide the selection of models and the tuning of hyper-parameters. For the evaluation of the proposed model's performance, different evaluation metrics are applied to the output on the test sets. Those evaluation metrics are presented mathematically below.

$$Precision \quad Pr = \frac{TP}{TP + FP} \quad (1)$$

$$Recall \quad Re = \frac{TP}{TP + FN} \quad (2)$$

$$F1\text{-score} \quad F1 = \frac{2TP}{2TP + FP + FN} \quad (3)$$

$$Accuracy \quad ACC = \frac{TP + TN}{TP + TN + FP + FN} \quad (4)$$

$$Misclassification\ Rate \quad MR = \frac{Incorrect\ predictions}{total\ predictions} \quad (5)$$

$$Macro\text{-}F1 \quad mF1 = \frac{\sum_{i=1}^{n} F1_i}{n} \quad (6)$$

$$Weighted\text{-}F1 \quad wF1 = \sum_{i=1}^{n} w_i * F1_i \quad (7)$$

Here, $TP$ is true positive, $TN$ is true negative, $FP$ is false positive, $FN$ is false negative, $n$ is the total number of classes, $F1_i$ is the F1-score for the $i$-th class, and $w_i = (No.of\ samples\ in\ class\ i)/(Total\ no.of\ samples)$.

### 3.4 Implementation of the Proposed CNN Model

This study introduces a CNN model for skin lesion classification based on the number of contained lesion colors. The model architecture consists of 19 layers, including convolutional layers, normalization layers, activation functions, pooling layers, and fully connected layers, culminating in a SoftMax output for classification. The primary focus of the research is to evaluate the impact of different configuration strategies on the model's performance by incorporating skip connections, also known as residual blocks [36], at various stages of the network. These experiments were conducted using three widely recognized skin lesion datasets and one combined dataset: PH2, Med-Node, PH2 and Med-Node (P+M), and ISIC2016. The proposed model consists of a CNN with a total of 19 layers, inspired by LeCun et al. [37] and Rasel et al. [38].

The initial design of the proposed model began with a 31-layer architecture. Through ablation studies, layers that contributed minimally to performance were systematically removed, resulting in an optimized 19-layer model that effectively classifies skin lesions based on the number of lesion colors. This reduction involved removing layers 17 to 28 due to limited training data availability, demonstrating that a more compact network could achieve comparable results. The final 19-layer configuration includes essential components such as input, convolutional, normalization, activation, max pooling, fully connected, SoftMax, and class output layers. Subsequently, hyperparameter tuning was performed to further enhance the model's performance, with the optimal settings documented in **Table 5**. Additionally, three extra layers were added during this tuning process to maximize the network's efficacy.

The model processes input images of size 256x256x3, representing RGB color images of skin lesions. The architecture begins with several convolutional layers, each utilizing a 5x5 kernel size with 32 filters, designed to extract relevant features for lesion classification. To maintain the spatial dimensions of the feature maps, padding is applied in the "same" mode, and a dilation factor of 2x2 is used to expand the receptive field without increasing the number of parameters. After each convolutional layer, a normalization layer is employed, followed by a ReLU activation function to introduce non-linearity into the model, which is essential for stabilizing and accelerating the learning process. Max pooling layers with a 5x5 kernel size and a stride of 2x2 are then applied to reduce the spatial dimensions of the feature maps, enabling the model to concentrate on the most prominent features.

A key feature of the model is the inclusion of skip connections, or residual blocks, at various stages (specifically at layers 10, 15, and 16). These skip connections create direct pathways between earlier and later layers, allowing the model to bypass one or more layers and directly add the output from a previous layer to a later one. This strategy mitigates the vanishing gradient problem, facilitates the training of deeper networks, and helps in learning more complex patterns by combining information from different stages of the network. The model culminates in a fully connected layer with 1,572,867 neurons, which is followed by a SoftMax output layer that produces a probability distribution across the three classes (A, B, and C), ultimately determining the classification outcome. This architecture is designed in **Table 5** to optimize the model's ability to generalize across different datasets and enhance its performance in clinical applications, particularly for the early detection and classification of skin lesions, which could significantly aid in the diagnosis of conditions such as melanoma.

**Table 5**
The description of the proposed CNN model for skin lesion classification.

| No. | Layer Type | Kernel Size* | No. of Filters/Neurons* | No. of Learnable | Dilation Factor* | Padding* | Stride* |
|---|---|---|---|---|---|---|---|
| 1 | Image/Input (256x256x3) | - | - | 0 | - | - | - |
| 2 | Convolution | 5x5 | 32 | 2432 | 2x2 | same | 1x1 |
| 3 | Normalization | - | - | 64 | - | - | - |
| 4 | ReLU | - | 32 | 0 | - | - | - |
| 5 | Max Pooling | 5x5 | - | 0 | - | same | 2x2 |
| 6 | Convolution | 5x5 | 32 | 25632 | 2x2 | same | 2x2 |
| 7 | Normalization | - | - | 64 | - | - | - |
| 8 | ReLU | - | 32 | 0 | - | - | - |
| 9 | Max Pooling | 5x5 | - | 0 | - | same | 2x2 |
| 10 | Addition** | | | 0 | | | |
| 11 | Convolution | 5x5 | 32 | 25632 | 2x2 | same | 2x2 |
| 12 | Normalization | - | - | 64 | - | - | - |
| 13 | ReLU | - | 32 | 0 | - | - | - |
| 14 | Max Pooling | 5x5 | - | 0 | - | same | 2x2 |
| 15 | Addition** | | | 0 | | | |
| 16 | Addition** | | | 0 | | | |
| 17 | Convolution | 5x5 | 32 | 25632 | 2x2 | same | 2x2 |
| 18 | Normalization | - | - | 64 | - | - | - |
| 19 | ReLU | - | 32 | 0 | - | - | - |
| 20 | Fully connected | - | - | 1572867 | - | - | - |
| 21 | Softmax | - | - | 0 | - | - | - |
| 22 | Class Output | - | - | 0 | - | - | - |

*The parameters with (\*) were selected by the "Grid-Search" method to optimize the network's performance. The layers with (\*\*) were connected by using skip-connections.*

The research is divided into four experiments to assess the effect of the addition layers (skip connections) on the model's performance. Experiment 1: The model is trained without any skip connections in (a) of **Fig. 3**, experiment 2: A skip connection is introduced at layer 10 in (b) of **Fig. 3**, experiment 3: Skip connections are incorporated at layers 10 and 15 in (c) of **Fig. 3**, and experiment 4: All three skip connections at layers 10, 15, and 16 are included in (d) of **Fig. 3**. The model is trained and evaluated on the PH2, Med-Node, P+M, and ISIC datasets for each experiment, with a focus on comparing classification accuracy, precision, recall, and F1-score. This comparative analysis provides insights into how the use of skip connections at different stages of the network influences the classification of skin lesions based on color, contributing to the advancement of automated diagnostic tools in dermatology. To gain deeper insights into the internal workings of the proposed CNN model, we utilized the DeepDream [39] algorithm to visualize the features extracted by specific layers, namely layers 2, 6, 16, and 17. DeepDream is a process that enhances and amplifies the patterns that a neural network has learned, allowing us to observe the types of features that each layer focuses on during the classification task. By applying this technique to selected convolutional layers in **Fig. 4**, we can better understand how the network interprets and differentiates skin lesions based on color.

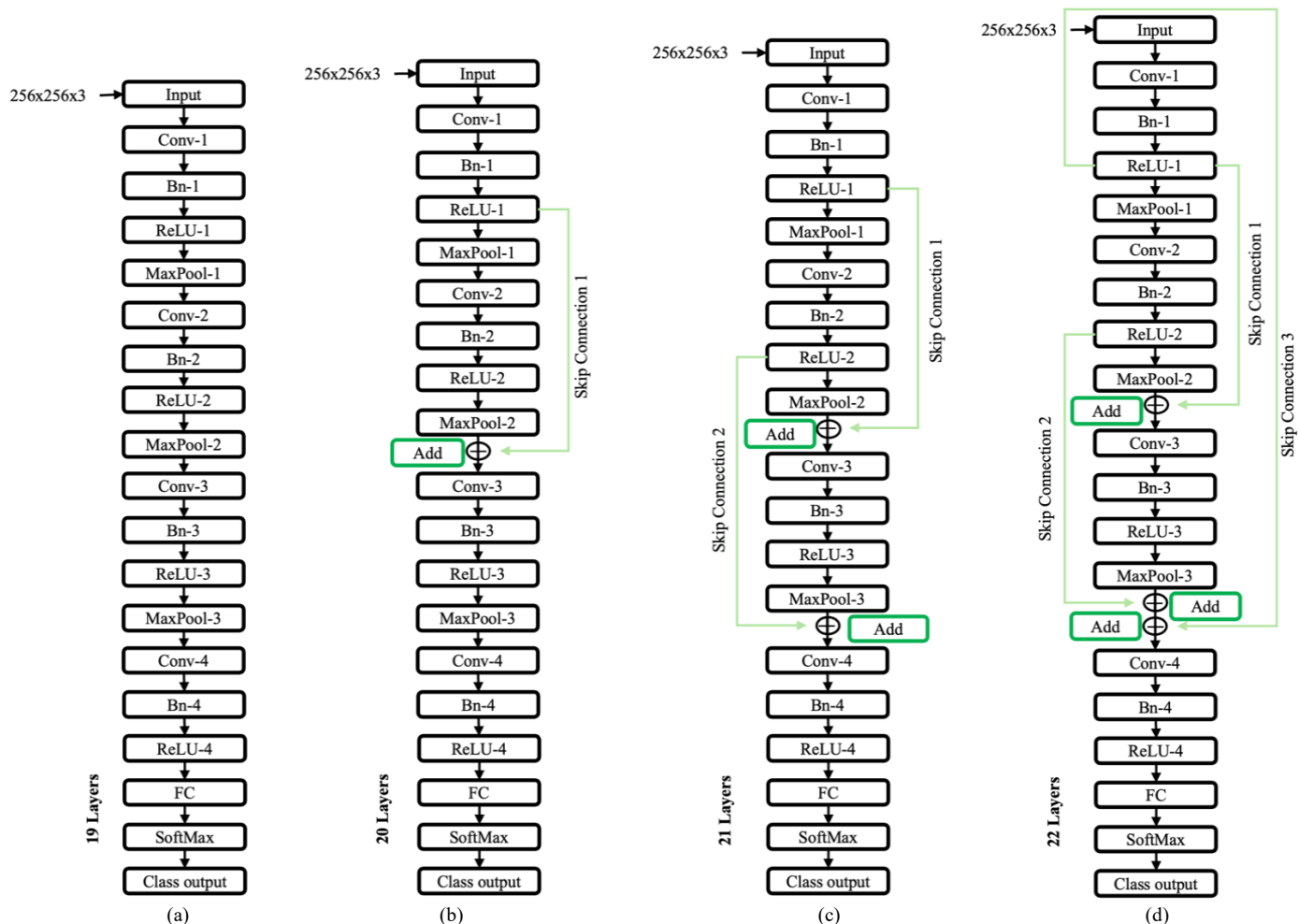

**Fig. 3.** Proposed model's architecture.

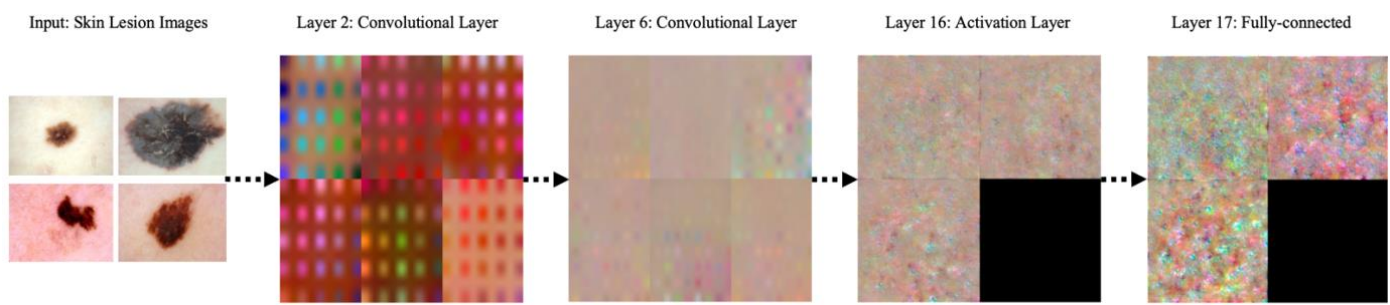

**Fig. 4.** Visualization of extracted features on different layers by using the DeepDream algorithm.

Layer 2, one of the initial convolutional layers, is responsible for capturing low-level features such as edges and basic color gradients. The DeepDream visualizations for this layer reveal simple patterns that form the foundational elements of the lesion images. Moving deeper into the network, Layer 6 extracts more complex features, where the visualizations begin to show combinations of basic shapes and textures that the model has learned to associate with specific color patterns in the lesions. Layer 16, which is positioned after multiple convolutional and pooling layers and includes two skip connections, displays even more intricate and abstract patterns in its DeepDream visualizations. This layer captures higher-level features that are crucial for distinguishing between different categories of lesions. The complexity of the patterns suggests that the network is identifying distinct color combinations and structural elements that are characteristic of skin lesion types. Finally, Layer 17, another convolutional layer following a skip connection, refines these high-level features further. The visualizations here highlight how the network amplifies specific aspects of the lesion, focusing on the most significant features that contribute to the final classification decision. These amplified patterns indicate that the model has developed a subtlety understanding of the lesion images, incorporating the intricate details necessary for accurate classification. By analyzing these DeepDream visualizations, we can confirm that the model progressively learns more complex and abstract features as the data moves through the layers. This visualization provides a deeper understanding of the model's feature extraction process as well as validates the effectiveness of the skip connections in enhancing the network's ability to capture and utilize critical information for classification.

**4. Results**

The PH2 dataset served as the reference standard for classifying lesions in the Med-Node and ISIC2016 datasets based on the number of colors present. An analysis of the Med-Node dataset revealed the following frequency of colors in the 170 lesion images: light brown appeared 113 times, dark brown 93 times, red 56 times, blue-gray 25 times, white 36 times, and black 45 times. In the ISIC2016 dataset, light brown was found 200 times, dark brown 103 times, red 196 times, blue-gray 90 times, white 79 times, and black 110 times in the 379 lesion images. Following the lesion color analysis detailed in Section 3.2, the categorization of these skin lesion images is summarized in **Table 6**, which categorizes the images from the PH2, Med-Node, PH2 and Med-Node (P+M), and ISIC2016 datasets based on the number of colors present.

**Table 6**
Categorization of lesion images in the PH2, Med-Node, PH2 and Med-Node (P+M), and ISIC2016 datasets.

| Datasets | Total Number | Any One Color | Any Two Colors | Any Three Colors | Any Four Colors | Any Five Colors | Any Six Colors |
|---|---|---|---|---|---|---|---|
| PH2 | 200 | 54 | 107 | 23 | 13 | 3 | - |
| Med-Node | 170 | 64 | 53 | 30 | 13 | 4 | 6 |
| P+M | 370 | 118 | 160 | 53 | 26 | 7 | 6 |
| ISIC2016 | 379 | 119 | 142 | 100 | 16 | 1 | 1 |

To further analyze the skin lesion datasets, we divided each dataset into three distinct classes based on the number of colors present in the lesions: Class A, Class B, and Class C. Class A consists of lesions with only one color, Class B includes lesions with any two colors, and Class C encompasses lesions with three or more colors (including 4, 5, and 6 colors). The rationale behind this classification is the relatively low number of skin lesions that exhibit more than three colors. By grouping lesions with three or more colors into a single class (Class C), we aim to create a more balanced distribution across the classes, facilitating a more effective and meaningful analysis. The classification of lesion images into these three classes for each dataset is summarized in **Table 7**.

**Table 7**
Classification of lesion images into Class A, B, and C across the PH2, Med-Node, P+M, and ISIC2016 datasets.

| Datasets | Total Number | Class A (One Color) | Class B (Two Colors) | Class C (Three or More Colors) |
|---|---|---|---|---|
| PH2 | 200 | 54 | 107 | 39 |
| Med-Node | 170 | 64 | 53 | 53 |
| P+M | 370 | 118 | 160 | 92 |
| ISIC2016 | 379 | 119 | 142 | 118 |

This classification scheme allows for a better understanding of the distribution of lesion colors across the datasets. That provides a solid foundation for subsequent analyses and model training, where distinguishing between different color patterns is critical for accurate skin lesion classification. We then trained our proposed CNN models on the four datasets—PH2, Med-Node, P+M, and ISIC2016—using a training-to-testing data split of 80% and 20%, respectively. The 20% testing data was kept entirely unknown to the models during training to ensure an unbiased evaluation. As detailed in Section 3.4, four different models were implemented, each with varying configurations of layers, corresponding to the four experiments conducted: Experiment 1, Experiment 2, Experiment 3, and Experiment 4. These experiments were designed to assess the impact of different architectural modifications, such as the inclusion of skip connections, on the model's performance. The results of these experiments are presented in **Fig. 5** and **Fig. 6**. Using equations 1 to 7, the models' performance metrics are calculated and systematically reported in **Tables 8**, **9**, **10**, and **11**. Each table presents the outcomes of all experiments on an individual dataset, allowing for a comprehensive comparison of how the different configurations influenced classification accuracy, precision, recall, and F1-score. This structured approach ensures a clear understanding of the effectiveness of each model variation and provides valuable insights into the optimal architecture for skin lesion classification based on color patterns.

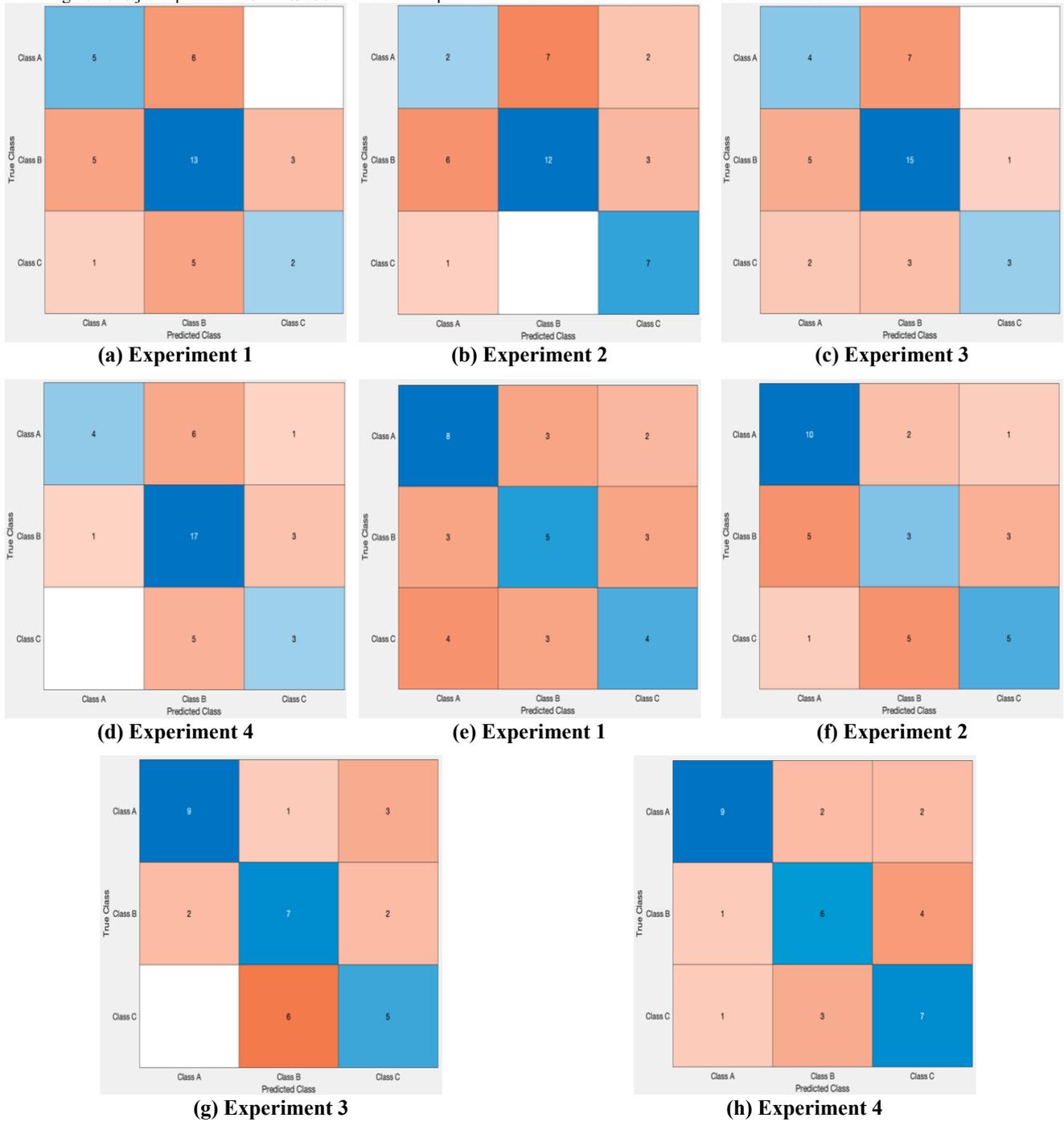

**Fig. 5.** The confusion matrices of all experiments 1 to 4 on PH2 (**a** to **d**), and Med-Node (**e** to **h**) datasets. Different evaluation metrics are applied to these confusion matrices as reported in **Tables 8** and **9**.

**Table 8**
Outputs of the proposed models on the PH2 dataset (training set 80% and testing set 20%).

| Model | Class A | | | Class B | | | Class C | | | Overall | | | |
|---|---|---|---|---|---|---|---|---|---|---|---|---|---|
| | Pr | Re | F1 | Pr | Re | F1 | Pr | Re | F1 | ACC | MR | mF1 | wF1 |
| Experiment 1 | 45 | 45 | 45 | 62 | 54 | 58 | 25 | 40 | 31 | 50 | 50 | 45 | 51 |
| Experiment 2 | 18 | 22 | 20 | 57 | 63 | 60 | 88 | 58 | 70 | 53 | 48 | 50 | 54 |
| Experiment 3 | 36 | 36 | 36 | 71 | 60 | 65 | 38 | 75 | 50 | 55 | 45 | 51 | 56 |
| Experiment 4 | 36 | 80 | 50 | 81 | 61 | 69 | 38 | 43 | 40 | 60 | 40 | 53 | 62 |

**Table 9**
Outputs of the proposed models on the Med-Node dataset (training set 80% and testing set 20%).

| Model | Class A | | | Class B | | | Class C | | | Overall | | | |
|---|---|---|---|---|---|---|---|---|---|---|---|---|---|
| | Pr | Re | F1 | Pr | Re | F1 | Pr | Re | F1 | ACC | MR | mF1 | wF1 |
| Experiment 1 | 62 | 53 | 57 | 45 | 45 | 45 | 36 | 44 | 40 | 49 | 51 | 48 | 49 |
| Experiment 2 | 77 | 63 | 69 | 27 | 30 | 29 | 45 | 56 | 50 | 51 | 49 | 49 | 53 |
| Experiment 3 | 69 | 82 | 75 | 64 | 50 | 56 | 45 | 50 | 48 | 60 | 40 | 60 | 60 |

| Experiment 4 | 69 | 82 | 75 | 55 | 55 | 55 | 64 | 54 | 58 | 63 | 37 | 63 | 62 |

**Table 8** shows the outcomes of the proposed models on the PH2 dataset. Across the four experiments, we observe a general improvement in performance as the model architecture becomes more complex. Experiment 1, which does not include any skip connections, shows baseline performance, while Experiment 4, with the most comprehensive use of skip connections, achieves the highest F1-scores across most classes and the best overall accuracy (60%), macro F1-score (53%), and weighted F1-score (62%). **Table 9** presents the results for the Med-Node dataset. Like the PH2 dataset, there is a noticeable improvement in performance from all experiments 1 to 4. Experiment 4 outperforms the other models with a 75% F1-score in Class A, 55% in Class B, and 58% in Class C. This experiment also achieves the highest accuracy (63%), macro F1-score (63%), and weighted F1-score (62%), reflecting the model's enhanced ability to generalize across different lesion types when skip connections are incorporated.

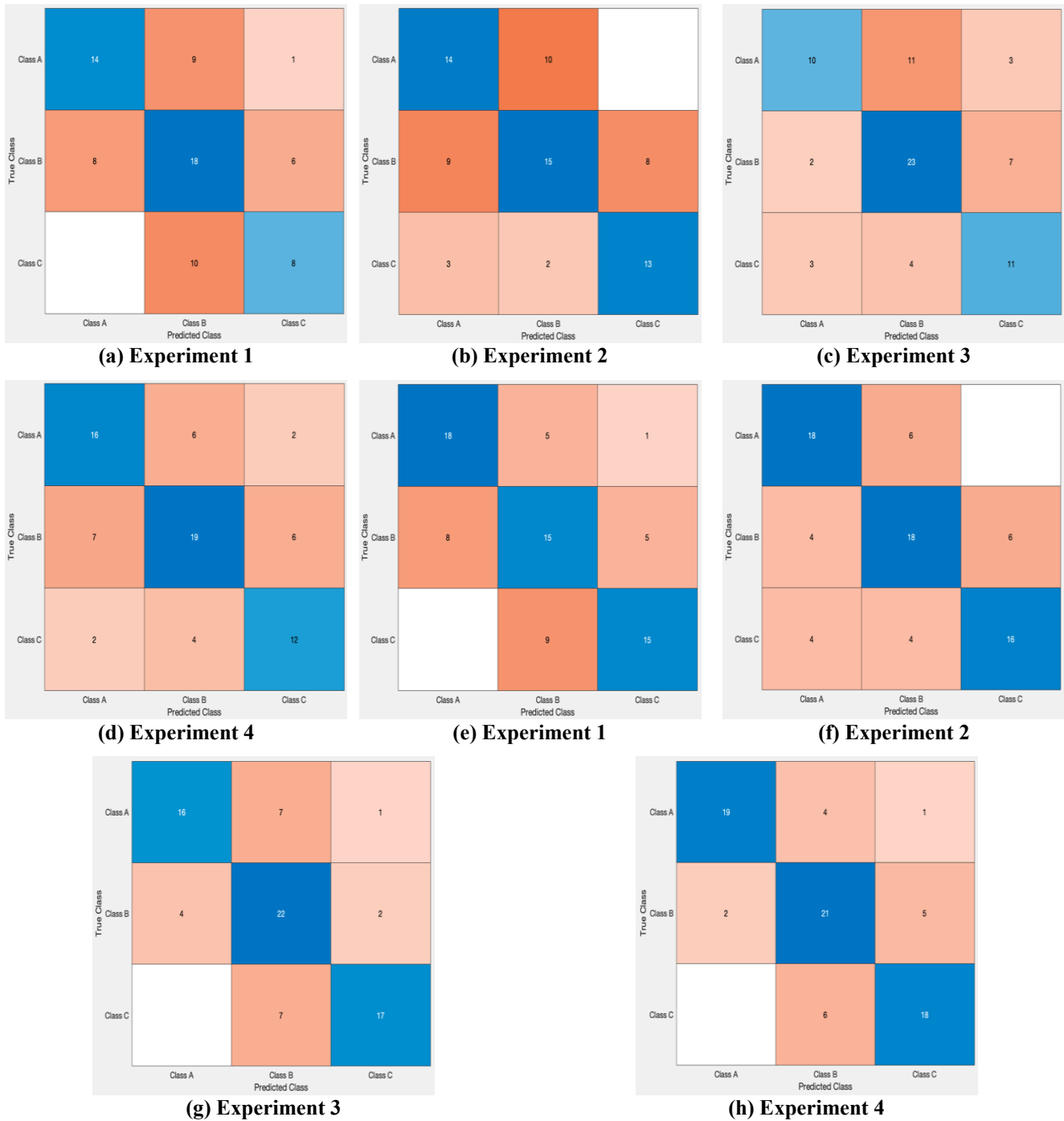

**Fig. 6.** The confusion matrices of all experiments 1 to 4 on P+M (a to d), and ISIC2016 (e to h) datasets. Different evaluation metrics are applied to these confusion matrices as reported in **Tables 10** and **11**.

**Table 10**
Outputs of the proposed models on the P+M dataset (training set 80% and testing set 20%).

| Model | Class A | | | Class B | | | Class C | | | Overall | | | |
|---|---|---|---|---|---|---|---|---|---|---|---|---|---|
| | Pr | Re | F1 | Pr | Re | F1 | Pr | Re | F1 | ACC | MR | mF1 | wF1 |
| Experiment 1 | 58 | 64 | 61 | 56 | 49 | 52 | 44 | 53 | 48 | 54 | 46 | 54 | 54 |
| Experiment 2 | 58 | 54 | 56 | 47 | 56 | 51 | 72 | 62 | 67 | 57 | 43 | 58 | 57 |

| Experiment 3 | 42 | 67 | 51 | 72 | 61 | 66 | 61 | 52 | 56 | 59 | 41 | 58 | 60 |
| Experiment 4 | 67 | 64 | 65 | 59 | 66 | 62 | 67 | 60 | 63 | 64 | 36 | 64 | 64 |

**Table 11**
Outputs of the proposed models on the ISIC2016 dataset (training set 80% and testing set 20%).

| Model | Class A | | | Class B | | | Class C | | | Overall | | | |
|---|---|---|---|---|---|---|---|---|---|---|---|---|---|
| | Pr | Re | F1 | Pr | Re | F1 | Pr | Re | F1 | ACC | MR | mF1 | wF1 |
| Experiment 1 | 75 | 69 | 72 | 54 | 52 | 53 | 63 | 71 | 67 | 63 | 37 | 64 | 63 |
| Experiment 2 | 75 | 69 | 72 | 64 | 64 | 64 | 67 | 73 | 70 | 68 | 32 | 69 | 68 |
| Experiment 3 | 67 | 80 | 73 | 79 | 61 | 69 | 71 | 85 | 77 | 72 | 28 | 73 | 72 |
| Experiment 4 | 79 | 90 | 84 | 75 | 68 | 71 | 75 | 75 | 75 | 76 | 24 | 77 | 76 |

**Table 10** reports the outputs for the P+M dataset, where the trend of increasing performance with more advanced model configurations continues. Experiment 4 again achieves the best performance across all metrics, with the highest overall accuracy of 64% and a weighted F1-score of 64%. This experiment demonstrates balanced performance across all three classes. These results indicate the effectiveness of the skip connections in handling the more complex lesion patterns present in this dataset. Finally, **Table 11** summarizes the results for the ISIC2016 dataset. The progression from Experiment 1 to Experiment 4 shows substantial gains, particularly in Class A, where Experiment 4 achieves an F1-score of 84%. This experiment also records the highest overall accuracy (76%), macro F1-score (77%), and weighted F1-score (76%), demonstrating the model's robustness in distinguishing between lesions with varying color complexities.

These tables collectively illustrate the positive impact of incorporating skip connections into the model architecture, particularly in enhancing the classification performance across different lesion types and color complexities in the datasets. The results underscore the importance of model architecture optimization in achieving higher accuracy and reliability in skin lesion classification tasks.

## 5. Discussion
### 5.1 Generalization of the proposed model
In **Tables 8** to **11**, the test and training sets were derived from the same datasets. While the test data was completely unknown during the training process, the fact that the test and training data originated from the same source may have influenced the evaluation results. To better demonstrate the generalization capabilities of the models, we conducted an additional experiment where the models were trained solely on the combined Med-Node and ISIC2016 datasets, excluding PH2 from the training process. We then tested the models on a randomly selected subset of 50% of the PH2 dataset, consisting of 100 skin lesion images. The performance of all four experiments on this independent PH2 test set is presented in **Fig.7** as confusion matrices and reported in **Table 12** after applying equations 1 to 7 on these matrices. This process was designed to evaluate the models' ability to generalize to new data that was not part of the training process, thereby offering a more robust assessment of their performance.

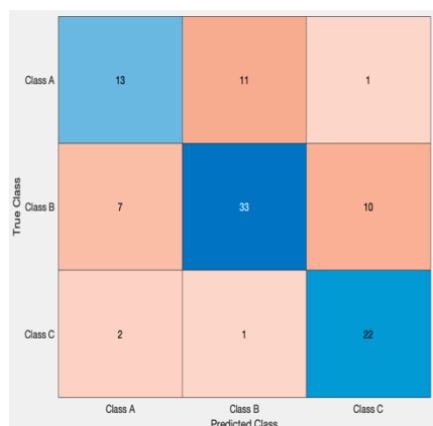
(a) Experiment 1

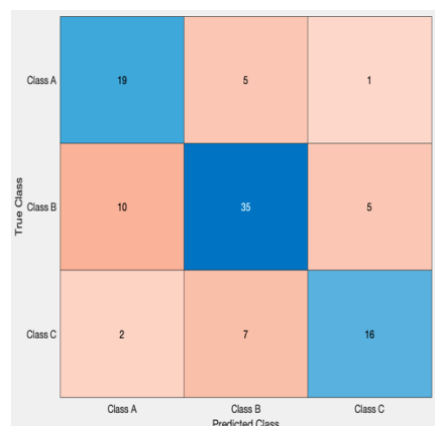
(b) Experiment 2

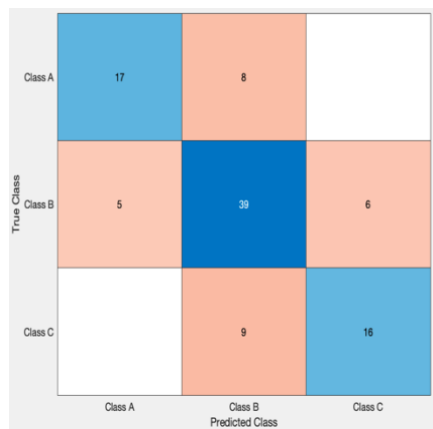

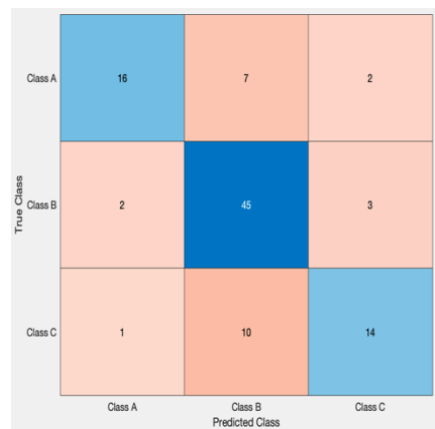

**(c) Experiment 3**            **(d) Experiment 4**

**Fig. 7.** The confusion matrices of the proposed models (in four experiments) on the test set (100 images) of the PH2 dataset when trained on the Med-Node and ISIC2016 datasets for the model generalization.

**Table 12**
Experimental results of the proposed models on the test set (100 images) of the PH2 dataset.

| Model | Class A | | | Class B | | | Class C | | | Overall | | | |
|---|---|---|---|---|---|---|---|---|---|---|---|---|---|
| | Pr | Re | F1 | Pr | Re | F1 | Pr | Re | F1 | ACC | MR | mF1 | wF1 |
| Experiment 1 | 52 | 59 | 55 | 66 | 73 | 69 | 88 | 67 | 76 | 68 | 32 | 67 | 68 |
| Experiment 2 | 76 | 61 | 68 | 70 | 74 | 72 | 64 | 73 | 68 | 70 | 30 | 69 | 70 |
| Experiment 3 | 68 | 77 | 72 | 78 | 70 | 74 | 64 | 73 | 68 | 72 | 28 | 71 | 72 |
| Experiment 4 | 64 | 84 | 73 | 90 | 73 | 80 | 56 | 74 | 64 | 75 | 25 | 72 | 76 |

**Table 12** shows the performance metrics for Class A (one color), Class B (two colors), and Class C (three or more colors), along with overall accuracy (ACC), misclassification rate (MR), macro F1-score (mF1), and weighted F1-score (wF1) for the PH2 test set. The results indicate a clear progression in performance across the four experiments, with Experiment 4 achieving the highest scores in most metrics. Experiment 4 achieves an F1-score of 73% for Class A, 80% for Class B, and 64% for Class C. The macro F1-score (72%) and weighted F1-score (76%) reflect the robustness of this model configuration, indicating its ability to classify the skin lesions more effectively than the other experiments.

*5.2 Applying LIME to the proposed model*
After training the model, we applied LIME (Local Interpretable Model-agnostic Explanations) [40] to visualize the key features influencing the model's decisions. LIME offers valuable insights into the specific features the trained model relies on when classifying skin lesion images. In **Fig. 8**, the model classified the skin lesion in image (a) as Class C with a probability of 0.98, while assigning probabilities of 0.01 to both Class A and Class B. Similarly, the model classified the skin lesion in image (b) as Class A with a probability of 0.99 and a probability of 0.01 for Class B. In **Fig. 9**, LIME highlights different regions of images (a) and (b) by overlaying hues of red, brown, and yellow to indicate the critical areas that distinguish Class C in image (a) and Class A in the image (b) from other classes. Blue and green areas represent regions that are common across multiple classes, suggesting less distinctive features. **Fig. 10** further illustrates LIME's identification of the top four important features or regions that led to the classification of image (a) as Class C and image (b) as Class A. The key regions in the image (a) contain more than two colors, leading the model to categorize it as Class C, where lesions typically contain three or more colors. Conversely, the key regions in the image (b) contain fewer than two colors, which is why the model categorized this skin lesion as Class A, where lesions are characterized by a single color. The application of LIME to the trained model significantly enhances our understanding of the decision-making process in lesion classification, making the classification process clearer, more interpretable, and ultimately more robust.

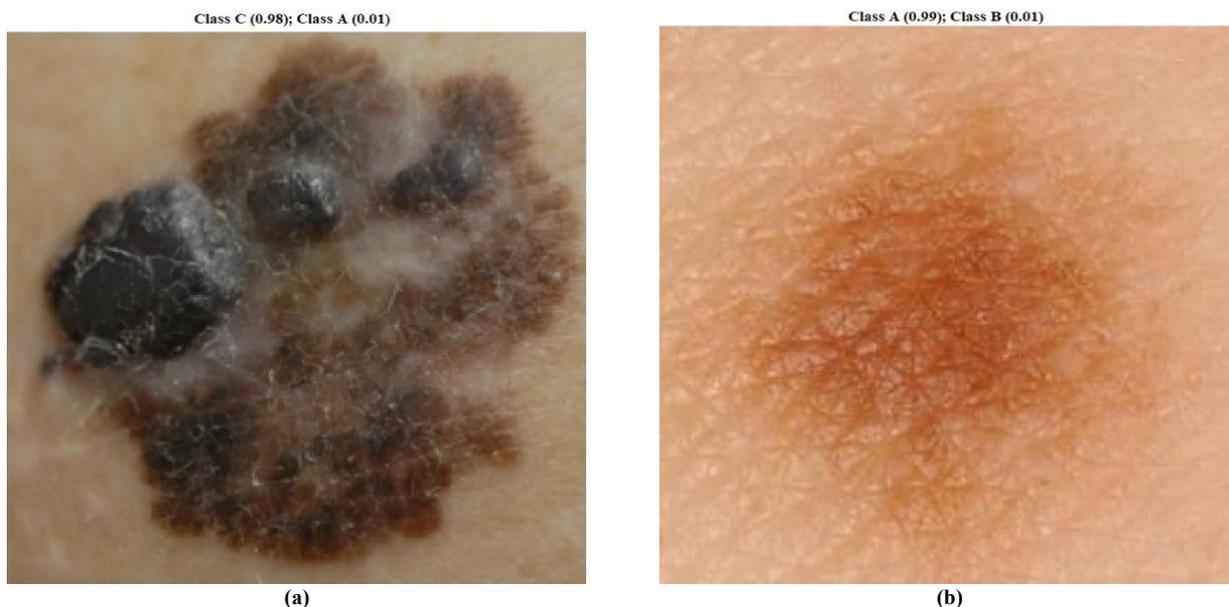

(a)            (b)
**Fig. 8.** The predicted class probability of the proposed model on two skin lesion images (a) 1685446.jpg, and (b) 21821.jpg.

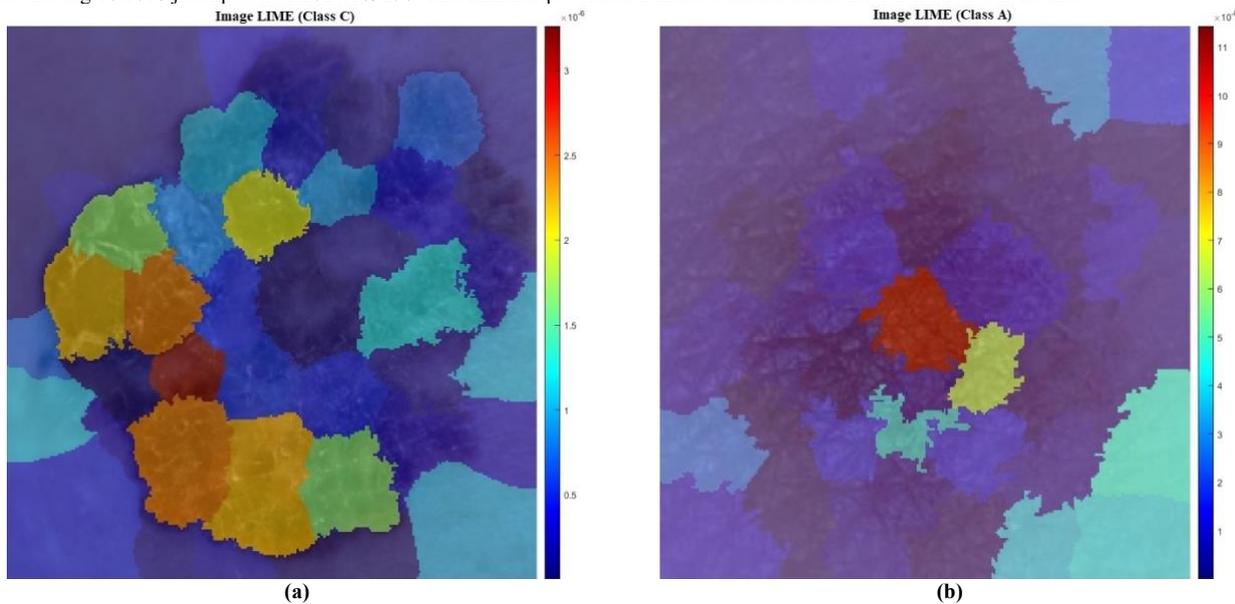

**Fig. 9.** The predicted class probability of the proposed model on two skin lesion images (a) 1685446.jpg, and (b) 21821.jpg.

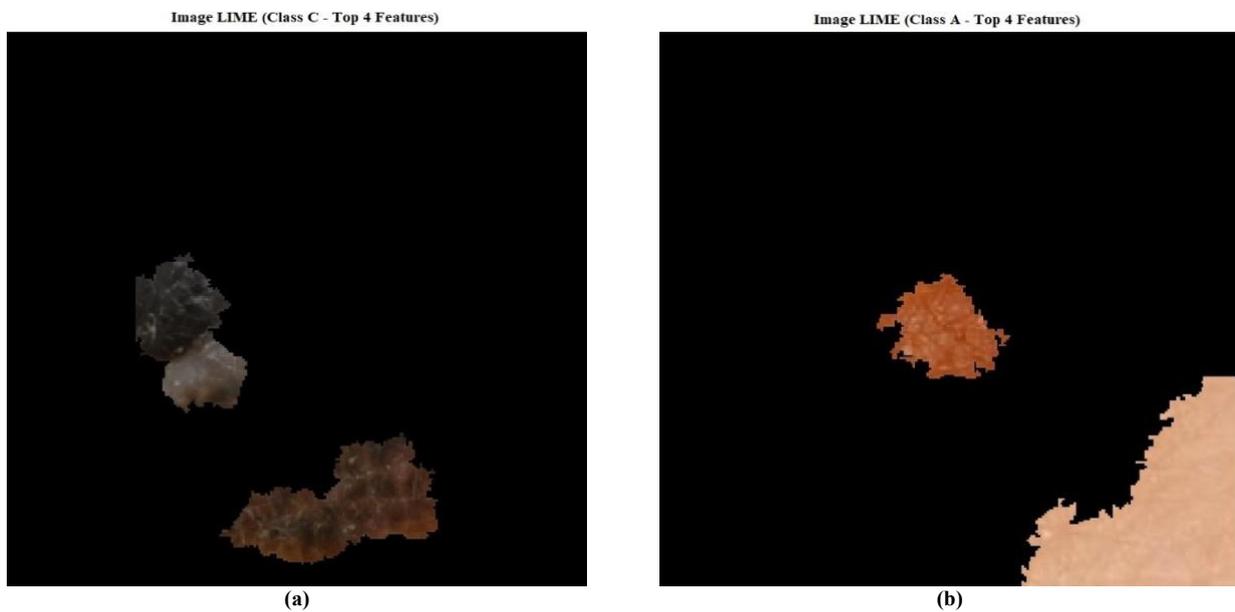

**Fig. 10.** The predicted class probability of the proposed model on two skin lesion images (a) 1685446.jpg, and (b) 21821.jpg.

These findings highlight the models' capacity to generalize effectively to unseen data, particularly when more complex architectures with skip connections (as in Experiment 4) are employed. This experiment underscores the importance of testing models on independent datasets to ensure their robustness and applicability in real-world clinical scenarios, where the data may differ from the training conditions.

*5.3 A Comparative Study*
To the best of our knowledge, no previous work has focused on classifying skin lesions based on the number of colors they contain. Most existing research has been dedicated to diagnosing specific conditions, such as differentiating between nevi and melanoma lesions. In this study, however, we exclusively analyze lesion colors, particularly those commonly found in melanoma lesions, and classify the lesions based on the number of colors present. This focus makes direct comparisons with previous work challenging, as no existing research has approached lesion classification in this manner. Given the lack of direct comparison opportunities, we utilized our proposed deep learning model for this novel task. To provide a benchmark, we adapted several state-of-the-art deep learning models traditionally used for diagnosing nevi or melanoma to classify lesions based on their color patterns. By retraining these models to recognize lesion classes according to the number of colors, we may compare the performance of our proposed model against these established models. This approach allows us to evaluate the effectiveness of our model in a broader context, even though the original purpose of the existing models differs from our specific classification task.

The first model was a combination of ResNet101 and SVM, where ResNet101 served as a feature extractor and SVM was trained on the extracted features [41]. The second model utilized a one-dimensional convolutional layer-based classifier that extracted 50 distinct features [42]. The third model was a customized CNN with 15 layers, employing various hyper-parameters to optimize performance [43]. To ensure a fair comparison, we implemented and trained all these models under the same conditions as our proposed model, using the Med-Node and ISIC2016 datasets together. The evaluation results of these experimental models on the PH2 test set are presented in **Table 13**.

Table 13
Comparing the performance of the proposed model with others on the test set (100 images) of the PH2 dataset.

| Model | Class A | | | Class B | | | Class C | | | Overall | | | |
|---|---|---|---|---|---|---|---|---|---|---|---|---|---|
| | Pr | Re | F1 | Pr | Re | F1 | Pr | Re | F1 | ACC | MR | mF1 | wF1 |
| Rasel et al. [41] | 76 | 59 | 67 | 68 | 85 | 76 | 76 | 68 | 72 | 72 | 28 | 71 | 72 |
| Kim et al. [42] | 72 | 64 | 68 | 68 | 77 | 72 | 72 | 64 | 68 | 70 | 30 | 69 | 70 |
| Shetty et al. [43] | 72 | 62 | 67 | 70 | 81 | 75 | 72 | 64 | 68 | 71 | 29 | 70 | 71 |
| **Proposed model** | **80** | **63** | **70** | **70** | **81** | **75** | **80** | **80** | **80** | **75** | **25** | **75** | **75** |

The proposed model (experiment 4) demonstrates superior performance across most metrics compared to the other models. Specifically, the proposed model achieves the highest F1-scores in all three classes: 70% for Class A, 75% for Class B, and 80% for Class C. The proposed model also records the highest overall accuracy (75%), lowest misclassification rate (25%), and highest macro and weighted F1-scores (75% each). These results indicate that the proposed model excels in accurately classifying skin lesions based on color patterns and outperforms the other state-of-the-art models in terms of overall reliability and robustness.

## 6. Conclusion

In this study, we developed a color detection and classification system for dermoscopic images to enhance melanoma diagnosis, focusing on critical color features such as black, dark brown, light brown, red, blue-gray, and white. Utilizing the PH2 dataset's ground truth and extracting RGB values from the literature, we analyzed each image from the Med-Node and ISIC2016 datasets to classify lesions into three classes: Class A (one color), Class B (two colors), and Class C (three or more colors). A 19-layer Convolutional Neural Network (CNN) was designed and trained using multiple skip connections of residual blocks to optimize performance, with the DeepDream algorithm ensuring network interpretability and LIME identifying significant features. The reported experiments included training the different architectures of CNN on PH2, Med-Node, a combined PH2 and Med-Node, and ISIC2016 datasets. The final model was tested on an independent set of PH2, demonstrating the model's robustness and generalizability. The proposed model, particularly with three skip connections, outperformed other approaches from the literature, providing valuable support to dermatologists in diagnosing melanoma and other color-associated skin conditions. However, this color-based study needs future investigation. The current classification approach is restricted to three classes based on the number of detected colors in lesions. Future studies should consider expanding this classification approach to include more classes, reflecting the diversity of dermoscopic colors. Besides, this model was trained using only three datasets, suggesting the need for more diverse dermoscopic datasets akin to PH2 to enhance the model's robustness. While the proposed 19-layer CNN model effectively classifies lesion colors, exploring deeper network architectures with more layers could further improve accuracy and efficiency. In conclusion, this research integrates pixel value analysis and CNN classification, contributing significantly to dermatology by potentially revolutionizing the diagnosis and treatment of skin conditions. Future studies should focus on expanding the classification technique, creating additional datasets, and exploring deeper network architectures to further advance this field.

Accepted Manuscript: This is the peer-reviewed version of the article accepted for publication in Computers in Biology and Medicine. Final published version: https://doi.org/10.1016/j.compbiomed.2024.109250. This manuscript version is made available under the CC BY-NC-ND license.11. Chen, J., Stanley, R.J., Moss, R.H., Van Stoecker, W., 2003. Colour analysis of skin lesion regions for melanoma discrimination in clinical images. Skin Research and Technology 9, 94–104. https://doi.org/10.1034/j.1600-0846.2003.00024.x
12. Ciudad-Blanco, C., Avilés-Izquierdo, J., Lázaro-Ochaita, P., & Suárez-Fernández, R. (2014). Dermoscopic Findings for the Early Detection of Melanoma: An Analysis of 200 Cases. *Actas Dermo-Sifiliográficas*, 105(7), 683-693. https://doi.org/10.1016/j.adengl.2014.07.015
13. Christos, P. J., Oliveria, S. A., Berwick, M., Guerry, D., Elder, D. E., Synnestvedt, M., Fine, J. A., Barnhill, R. L., & Halpern, A. C. (2000). Signs and symptoms of melanoma in older populations. *Journal of Clinical Epidemiology*, 53(10), 1044–1053. https://doi.org/10.1016/s0895-4356(00)00224-9
14. Mort, R. L., Jackson, I. J., & Patton, E. E. (2015). The melanocyte lineage in development and disease. *Development*, 142(7), 1387. https://doi.org/10.1242/dev.123729
15. MmFitzpatrick, T. B., Polano, M. K., & Suurmond, D. (1998). Color atlas and synopsis of clinical dermatology. *Journal of the American Academy of Dermatology*, 38(6), 1018–1019. https://doi.org/10.1016/s0190-9622(98)70192-6
16. Braun, R. P. (2002). Dermoscopy of Pigmented Seborrheic Keratosis. *Archives of Dermatology*, 138(12), 1556. https://doi.org/10.1001/archderm.138.12.1556
17. MmLiu, W., Dowling, J. P., Murray, W. K., McArthur, G. A., Thompson, J. F., Wolfe, R., & Kelly, J. W. (2006). Rate of Growth in Melanomas. *Archives of Dermatology*, 142(12). https://doi.org/10.1001/archderm.142.12.1551
18. Thawabteh, A. M., Jibreen, A., Karaman, D., Thawabteh, A., & Karaman, R. (2023). Skin Pigmentation Types, Causes and Treatment—A Review. *Molecules*, 28(12), 4839. https://doi.org/10.3390/molecules28124839
19. Guo, W., Wang, H., & Li, C. (2021). Signal pathways of melanoma and targeted therapy. *Signal Transduction and Targeted Therapy*, 6(1). https://doi.org/10.1038/s41392-021-00827-6
20. Wolner, Z.J., Yélamos, O., Liopyris, K., Rogers, T., Marchetti, M.A., Marghoob, A.A., 2017. Enhancing Skin Cancer Diagnosis with Dermoscopy. Dermatologic Clinics 35, 417–437. https://doi.org/10.1016/j.det.2017.06.003
21. Mendonca, T., Ferreira, P.M., Marques, J.S., Marcal, A.R.S., Rozeira, J., 2013. $PH^2$ - A dermoscopic image database for research and benchmarking. 2013 35th Annual International Conference of the IEEE Engineering in Medicine and Biology Society (EMBC). https://doi.org/10.1109/embc.2013.6610779
22. Liu, W., Dowling, J. P., Murray, W. K., McArthur, G. A., Thompson, J. F., Wolfe, R., & Kelly, J. W. (2006). Rate of Growth in Melanomas. *Archives of Dermatology*, 142(12). https://doi.org/10.1001/archderm.142.12.1551
23. Rezk, E., Eltorki, M., & El-Dakhakhni, W. (2022). Improving Skin Color Diversity in Cancer Detection: Deep Learning Approach. *JMIR Dermatology*, 5(3), e39143. https://doi.org/10.2196/39143
24. Stanley, R.J., Stoecker, W.V., Moss, R.H., 2007. A relative color approach to color discrimination for malignant melanoma detection in dermoscopy images. Skin Research and Technology 13, 62–72. https://doi.org/10.1111/j.1600-0846.2007.00192.x
25. Nezhadian, F.K., Rashidi, S., 2017. Melanoma skin cancer detection using color and new texture features. 2017 Artificial Intelligence and Signal Processing Conference (AISP). https://doi.org/10.1109/aisp.2017.8324108
26. Almubarak, H., Stanley, R., Stoecker, W., Moss, R., 2017. Fuzzy Color Clustering for Melanoma Diagnosis in Dermoscopy Images. Information 8, 89. https://doi.org/10.3390/info8030089
27. Rahman, M.A., Haque, M.T., Shahnaz, C., Fattah, S.A., Zhu, W.P., Ahmed, M.O., 2017. Skin lesions classification based on color plane-histogram-image quality analysis features extracted from digital images. 2017 IEEE 60th International Midwest Symposium on Circuits and Systems (MWSCAS). https://doi.org/10.1109/mwscas.2017.8053183
28. Sabbaghi Mahmouei, S., Aldeen, M., Stoecker, W.V., Garnavi, R., 2019. Biologically Inspired QuadTree Color Detection in Dermoscopy Images of Melanoma. IEEE Journal of Biomedical and Health Informatics 23, 570–577. https://doi.org/10.1109/jbhi.2018.2841428
29. Moldovanu, S., Michis, F. a. D., Biswas, K. C., Culea-Florescu, A., & Moraru, L. (2021). Skin Lesion Classification Based on Surface Fractal Dimensions and Statistical Color Cluster Features Using an Ensemble of Machine Learning Techniques. *Cancers*, 13(21), 5256. https://doi.org/10.3390/cancers13215256
30. Rasel, M. A., Kareem, S. A., Kwan, Z., Yong, S. S., & Obaidellah, U. (2024). Bluish veil detection and lesion classification using custom deep learnable layers with explainable artificial intelligence (XAI). *Computers in Biology and Medicine*, 178, 108758. https://doi.org/10.1016/j.compbiomed.2024.108758
31. Moldovanu, S., Miron, M., Rusu, C. G., Biswas, K. C., & Moraru, L. (2023). Refining skin lesions classification performance using geometric features of superpixels. *Scientific Reports*, 13(1). https://doi.org/10.1038/s41598-023-38706-5
32. Rout, R., & Parida, P. (2020). A novel method for melanocytic skin lesion extraction and analysis. *Journal of Discrete Mathematical Sciences and Cryptography*, 23(2), 461–473. https://doi.org/10.1080/09720529.2020.1728900
33. Sahoo, A. K., Parida, P., & Muralibabu, K. (2023). Hybrid deep neural network with clustering algorithms for effective gliomas segmentation. *International Journal of Systems Assurance Engineering and Management*, 15(3), 964–980. https://doi.org/10.1007/s13198-023-02183-w
34. Giotis, I., Molders, N., Land, S., Biehl, M., Jonkman, M.F., Petkov, N., 2015. MED-NODE: A computer-assisted melanoma diagnosis system using non-dermoscopic images. Expert Systems with Applications 42, 6578–6585. https://doi.org/10.1016/j.eswa.2015.04.034
35. Gutman, David; Codella, Noel C. F.; Celebi, Emre; Helba, Brian; Marchetti, Michael; Mishra, Nabin; Halpern, Allan. "Skin Lesion Analysis toward Melanoma Detection: A Challenge at the International Symposium on Biomedical Imaging (ISBI) 2016, hosted by the International Skin Imaging Collaboration (ISIC)". eprint arXiv:1605.01397. 2016.
36. Wang, Z., Yi, P., Jiang, K., Jiang, J., Han, Z., Lu, T., & Ma, J. (2019). Multi-Memory Convolutional Neural Network for Video Super-Resolution. *IEEE Transactions on Image Processing*, 28(5), 2530–2544. https://doi.org/10.1109/tip.2018.2887017
37. Lecun, Y., Bottou, L., Bengio, Y., & Haffner, P. (1998). Gradient-based learning applied to document recognition. *Proceedings of the IEEE*, 86(11), 2278–2324. https://doi.org/10.1109/5.726791
38. Rasel, M. A., Obaidellah, U. H., & Kareem, S. A. (2022). Convolutional Neural Network-Based Skin Lesion Classification With Variable Nonlinear Activation Functions. *IEEE Access*, 10, 83398–83414. https://doi.org/10.1109/access.2022.3196911
39. Szegedy, C., Liu, N. W., Jia, N. Y., Sermanet, P., Reed, S., Anguelov, D., Erhan, D., Vanhoucke, V., & Rabinovich, A. (2015c). *Going deeper with convolutions*. https://doi.org/10.1109/cvpr.2015.7298594
40. Kumarakulasinghe, N. B., Blomberg, T., Liu, J., Leao, A. S., & Papapetrou, P. (2020). *Evaluating Local Interpretable Model-Agnostic Explanations on Clinical Machine Learning Classification Models*. https://doi.org/10.1109/cbms49503.2020.00009
41. Rasel, M. A., Kareem, S. A., Kwan, Z., Faheem, N. a. a., Han, W. H., Choong, R. K. J., Yong, S. S., & Obaidellah, U. (2024). Asymmetric lesion detection with geometric patterns and CNN-SVM classification. *Computers in Biology and Medicine*, 179, 108851. https://doi.org/10.1016/j.compbiomed.2024.108851
42. Kim, C., Jang, M., Han, Y., Hong, Y., Lee, W., 2023. Skin Lesion Classification Using Hybrid Convolutional Neural Network with Edge, Color, and Texture Information. Applied Sciences 13, 5497. https://doi.org/10.3390/app13095497
43. Shetty, B., Fernandes, R., Rodrigues, A.P., Chengoden, R., Bhattacharya, S., Lakshmanna, K., 2022. Skin lesion classification of dermoscopic images using machine learning and convolutional neural network. Scientific Reports 12. https://doi.org/10.1038/s41598-022-22644-9